\documentclass[aps,prb,twocolumn,groupedaddress,showpacs,fleqn,floatfix]{revtex4}

\usepackage{amsmath}
\usepackage{epsfig}
\usepackage{color}

\newcommand{\figa} {
\begin{figure*}
\epsfig{file = 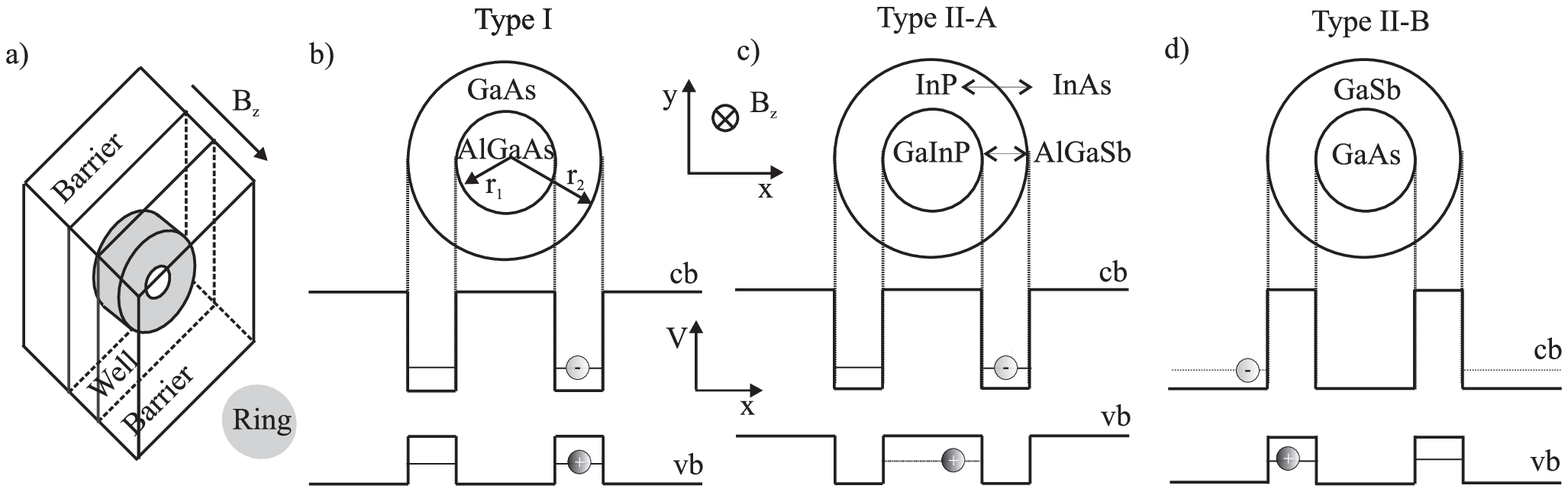, width = 16cm, angle = -0}
\caption{Schematic drawing of the investigated nanoring embedded in the quantum well (a) and the geometry within the $x$-$y$-plane: (b) type I, (c) type II-A, and (d) type II-B band alignment (see text). The magnetic field is directed along the growth direction $z$. Specific electron and hole positions (including strain) are visualized.}
\label{figa}
\end{figure*}
}

\newcommand{\figb}
{
\begin{figure}[th]
\begin{center}
\epsfig{file = 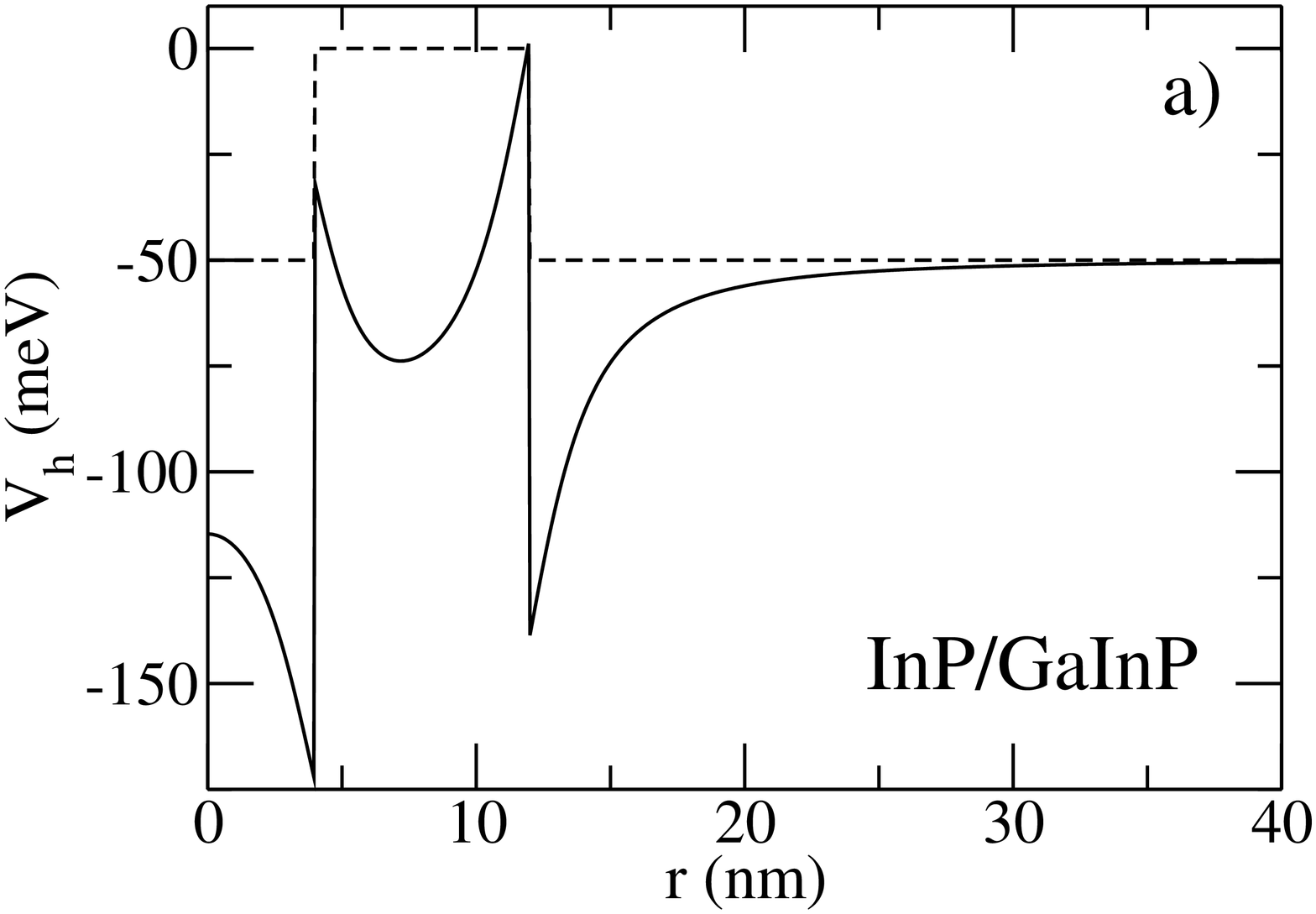, width = 5.9cm, angle = -90}
\epsfig{file = 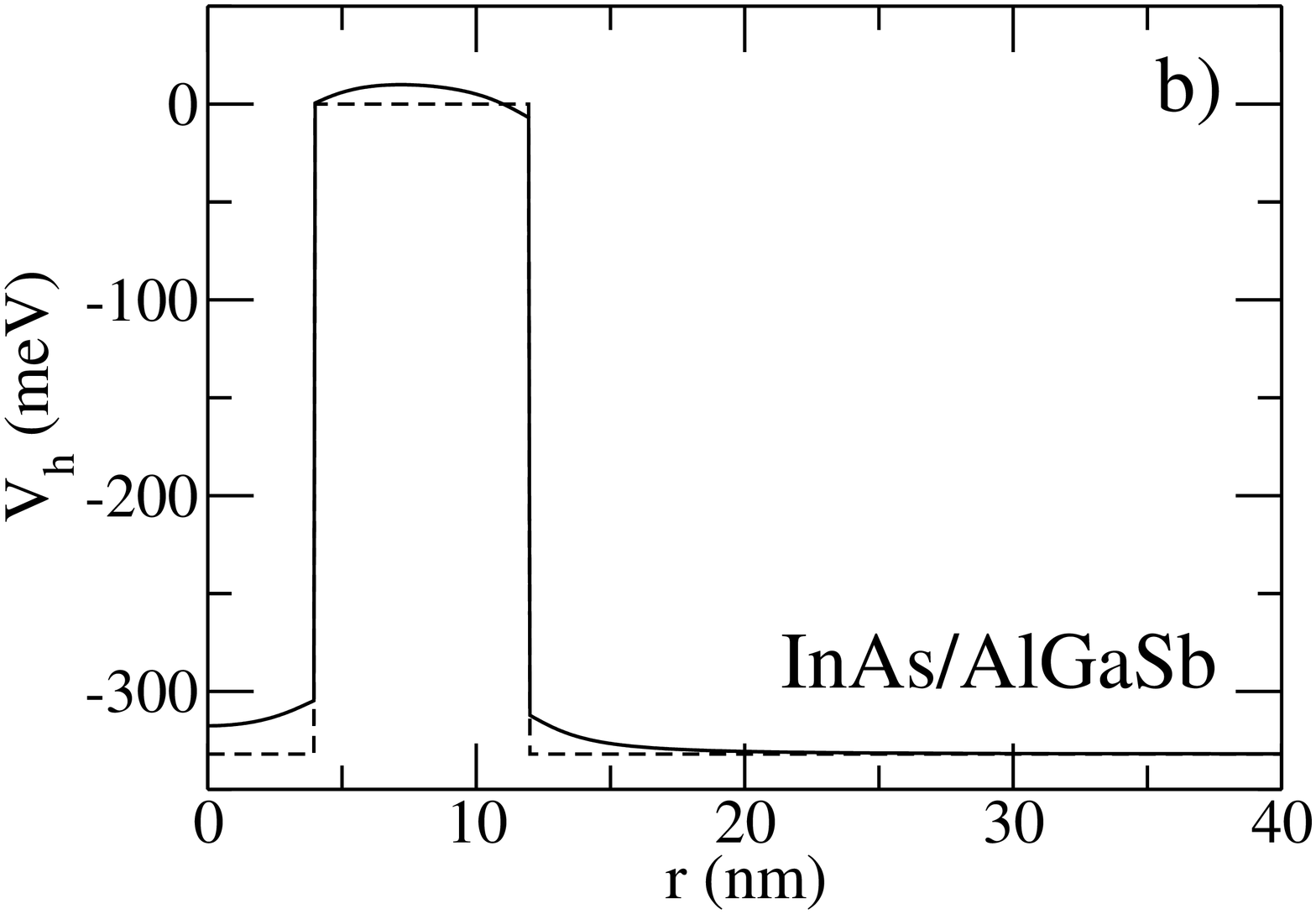, width = 5.9cm, angle = -90}
\caption{The hole potential in the $x$-$y$-plane for $z = 0$ of the InP/GaInP (a) and InAs/AlGaSb (b) nanorings without (dashed) and with (solid) strain included.}
\label{figb}
\end{center}
\end{figure}
}

\newcommand{\figc}
{
\begin{figure*}[t]
\begin{center}
\epsfig{file = 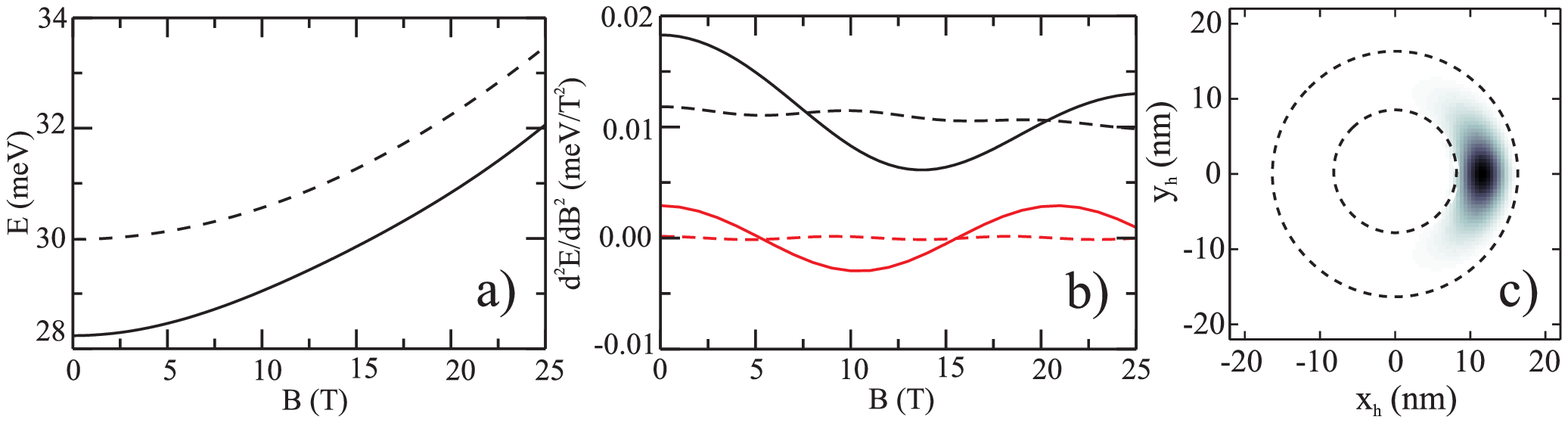, width = 16cm, angle = -0}
\caption{(Color online) The $B$-field dependence of lowest bright (ground) state energy (a) and its second derivative in type I GaAs/AlGaAs nanorings, solid~-~$r_1 = 4$~nm, $r_2 = 12$~nm, and dashed~-~$r_1 = 8$~nm, $r_2 = 16$~nm. The full calculation (black) is compared to results for infinitesimal narrow rings Eq. (\ref{SilvaModel2}) (red). The periods of the oscillations Eq.~(\ref{BP}) are $B_P = 20.8$~T (solid) and $B_P = 9.2$~T (dashed). Projected hole density $n^{(h)}$ according to Eq.~(\ref{densityh}) at $B = 0$~T (c). The ring boundaries are shown as dashed circles.}
\label{figc}
\end{center}
\end{figure*}
}

\newcommand{\figd}
{
\begin{figure*}[t]
\begin{center}
\epsfig{file = 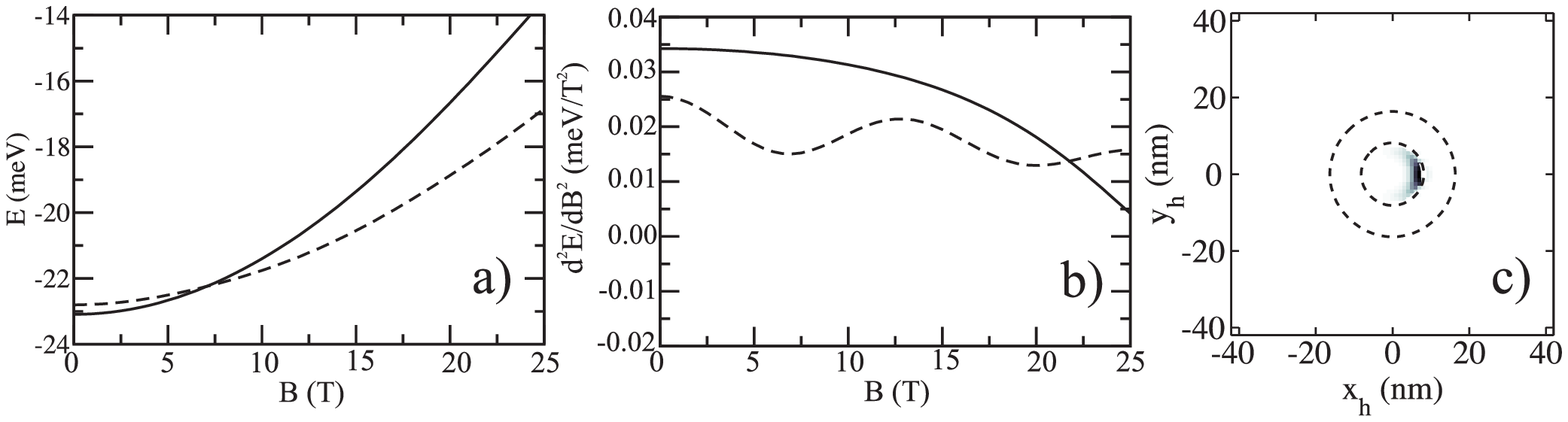, width = 16cm, angle = -0}
\caption{The $B$-field dependence of lowest bright state energy (a) and its second derivative (b). Solid~-~$r_1 = 4$~nm, $r_2 = 12$~nm, and dashed~-~$r_1 = 8$~nm, $r_2 = 16$~nm InP/GaInP nanoring. The solid line in (a) is shifted by 20 meV for comparison. Projected hole density $n^{(h)}$ according to Eq.~(\ref{densityh}) at $B = 0$~T (c). The ring boundaries are shown as dashed circles. The periods of the oscillations Eq.~(\ref{BP}) are $B_P = 40.9$~T (solid) and $B_P = 11.8$~T (dashed).}
\label{figd}
\end{center}
\end{figure*}
}

\newcommand{\figf}
{
\begin{figure*}[t]
\begin{center}
\epsfig{file = 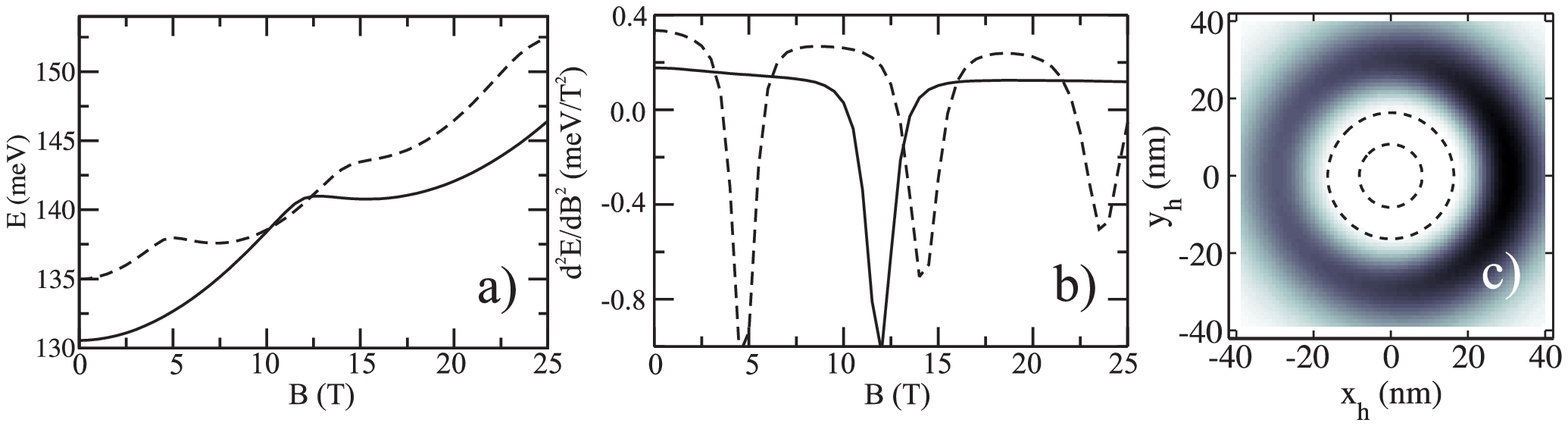, width = 16cm, angle = -0}
\caption{The $B$-field dependence of lowest bright state energy (a) and its second derivative (b). Solid~-~$r_1 = 4$~nm, $r_2 = 12$~nm, and dashed~-~$r_1 = 8$~nm, $r_2 = 16$~nm InAs/AlGaSb nanoring. Projected hole density $n^{(h)}$ according to Eq.~(\ref{densityh}) at $B = 0$~T (c). The ring boundaries are shown as dashed circles. The periods of the oscillations Eq.~(\ref{BP}) are $B_P = 19.9$~T (solid) and $B_P = 8.8$~T (dashed).}
\label{figf}
\end{center}
\end{figure*}
}

\newcommand{\figg}
{
\begin{figure*}[t]
\begin{center}
\epsfig{file = 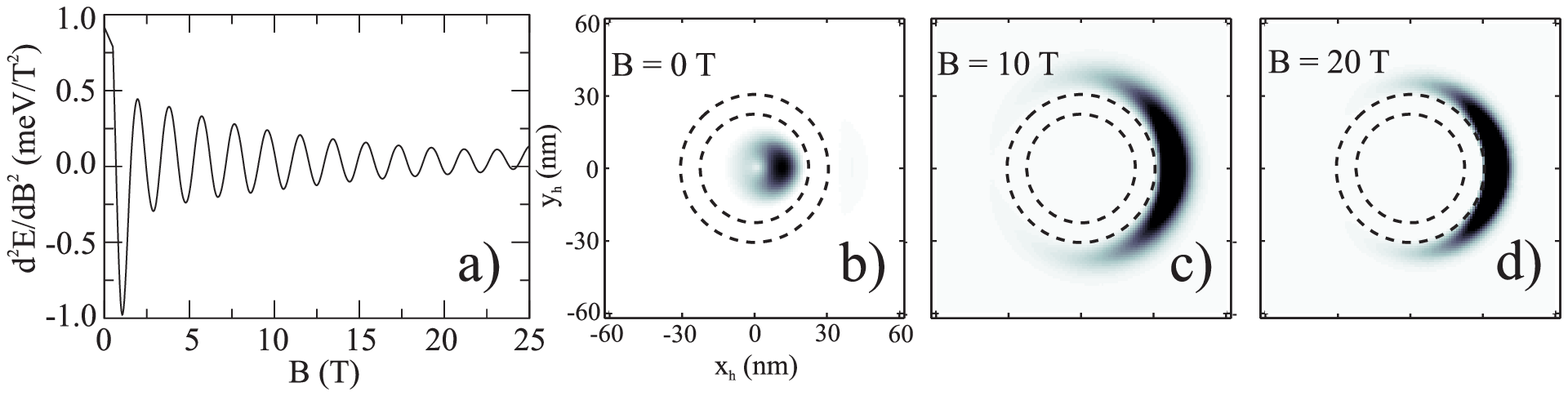, width = 16cm, angle = -0}
\caption{The second derivative with respect to the $B$-field of the lowest bright state energy (a) of the InAs/AlGaSb nanoring with radii~$r_1 = 22$~nm and $r_2 = 30$~nm. Projected hole density $n^{(h)}$ according to Eq.~(\ref{densityh}) at $B = 0$~T (b), $B = 10$~T (c), and $B=20$~T (d). The ring boundaries are shown as dashed circles. The period of the oscillations Eq.~(\ref{BP}) is $B_P = 1.9$~T. }
\label{figg}
\end{center}
\end{figure*}
}

\newcommand{\figh}
{
\begin{figure*}[t]
\begin{center}
\epsfig{file = 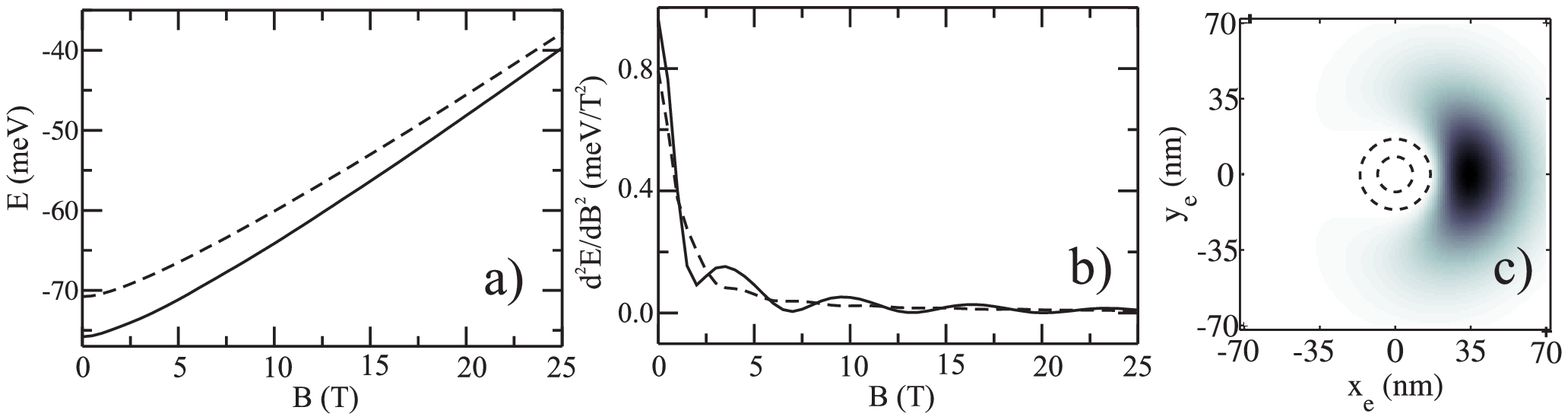, width = 16cm, angle = -0}
\caption{The $B$-field dependence of lowest bright state energy (a) and its second derivative (b). Solid~-~$r_1 = 4$~nm, $r_2 = 12$~nm, and dashed~-~$r_1 = 8$~nm, $r_2 = 16$~nm GaSb/GaAs nanoring. Projected electron density $n^{(e)}$ according to Eq.~(\ref{densitye}) at $B = 0$~T (c). The ring boundaries are shown as dashed circles. The periods of the oscillations Eq.~(\ref{BP}) are $B_P = 2.9$~T (solid) and $B_P = 1.5$~T (dashed).}
\label{figh}
\end{center}
\end{figure*}
}

\newcommand{\figi}
{
\begin{figure*}[t]
\begin{center}
\epsfig{file = 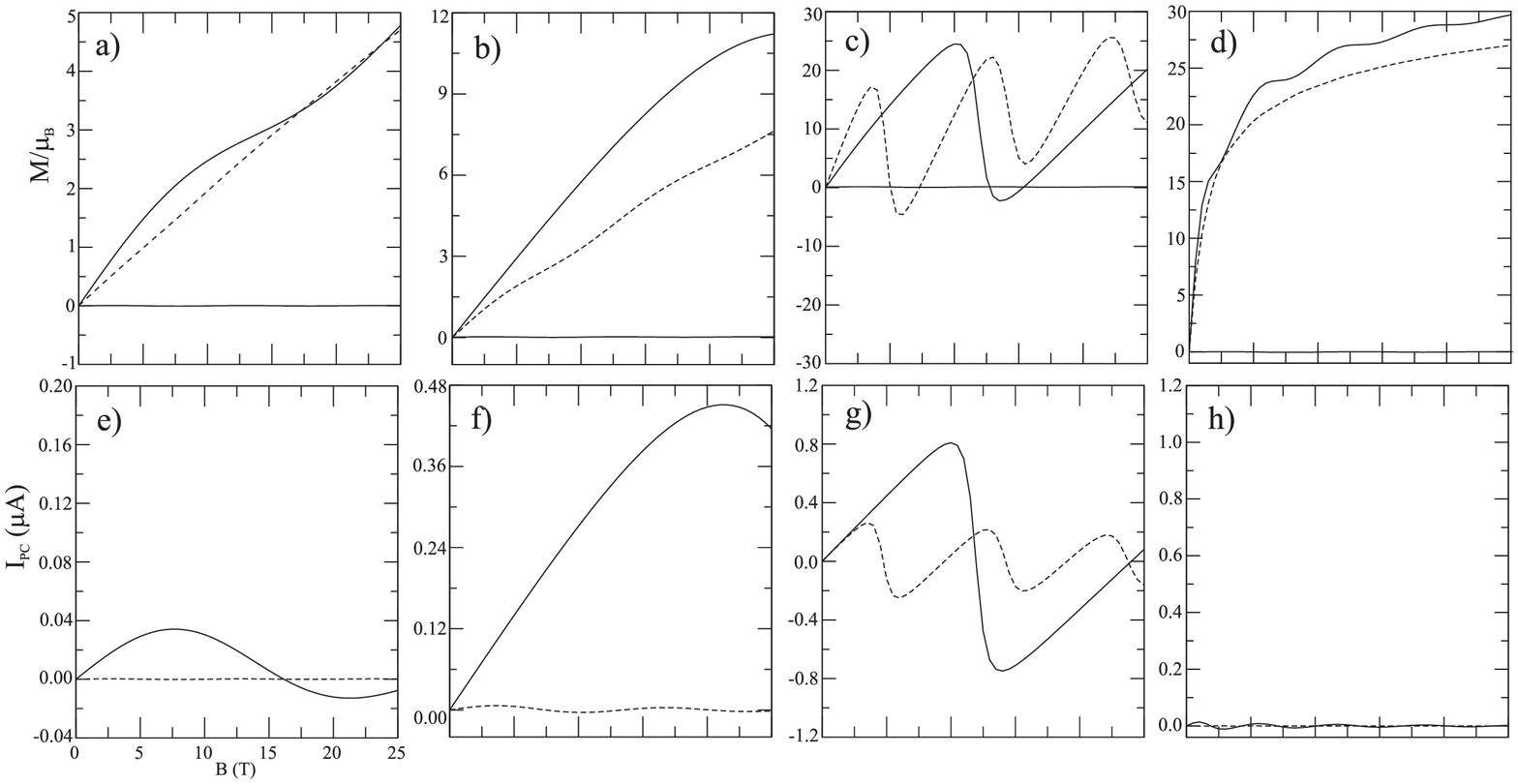, height = 16.cm, angle = 90}
\caption{Magnetization (a - d) and persistent current (e - h) and  of the nanorings with radii~$r_1 = 4$~nm, $r_2 = 12$~nm (solid) and ~$r_1 = 8$~nm, $r_2 = 16$~nm (dashed) of the lowest bright state. Materials: GaAs/AlGaAs (a, e), InP/GaInP (b, f), InAs/AlGaSb (c, g), and GaSb/GaAs (d, h). The scaling factor between the PC and the magnetization is held constant in all cases.}
\label{figi}
\end{center}
\end{figure*}
}

\newcommand{\figj}
{
\begin{figure*}[t]
\begin{center}
\epsfig{file = 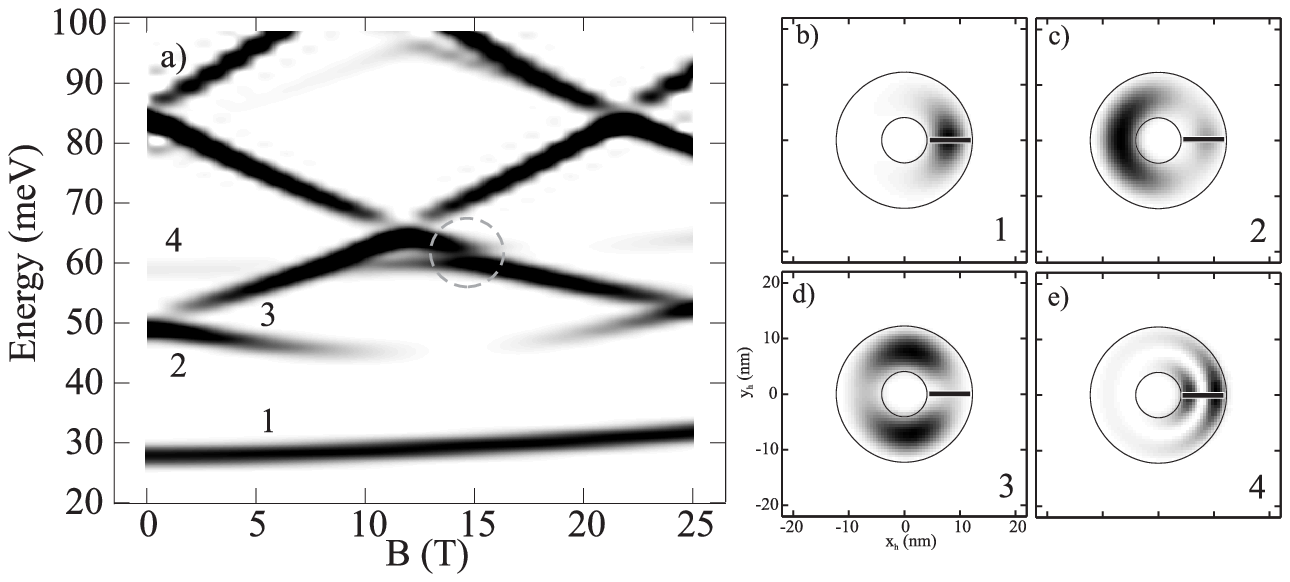, height = 16.cm, angle = 90}
\caption{Absorption spectrum (a) of a GaAs/AlGaAs nanoring with radii~$r_1 = 4$~nm and $r_2 = 12$~nm. The lines are Gauss broadened with $\sigma=1$~meV, and the oscillator strength (for the ground state divided by ten) is given in linear grey scale (step like features are artefacts of the interpolation). The circle focuses on the specific anti-crossing (see text). The correlated hole densities at $B=2$~T are given for the first four lowest bright states (b)-(e). The black rectangle indicates the fixed electron position.}
\label{figj}
\end{center}
\end{figure*}
}

\newcommand{\figk}
{
\begin{figure}[th]
\begin{center}
\epsfig{file = 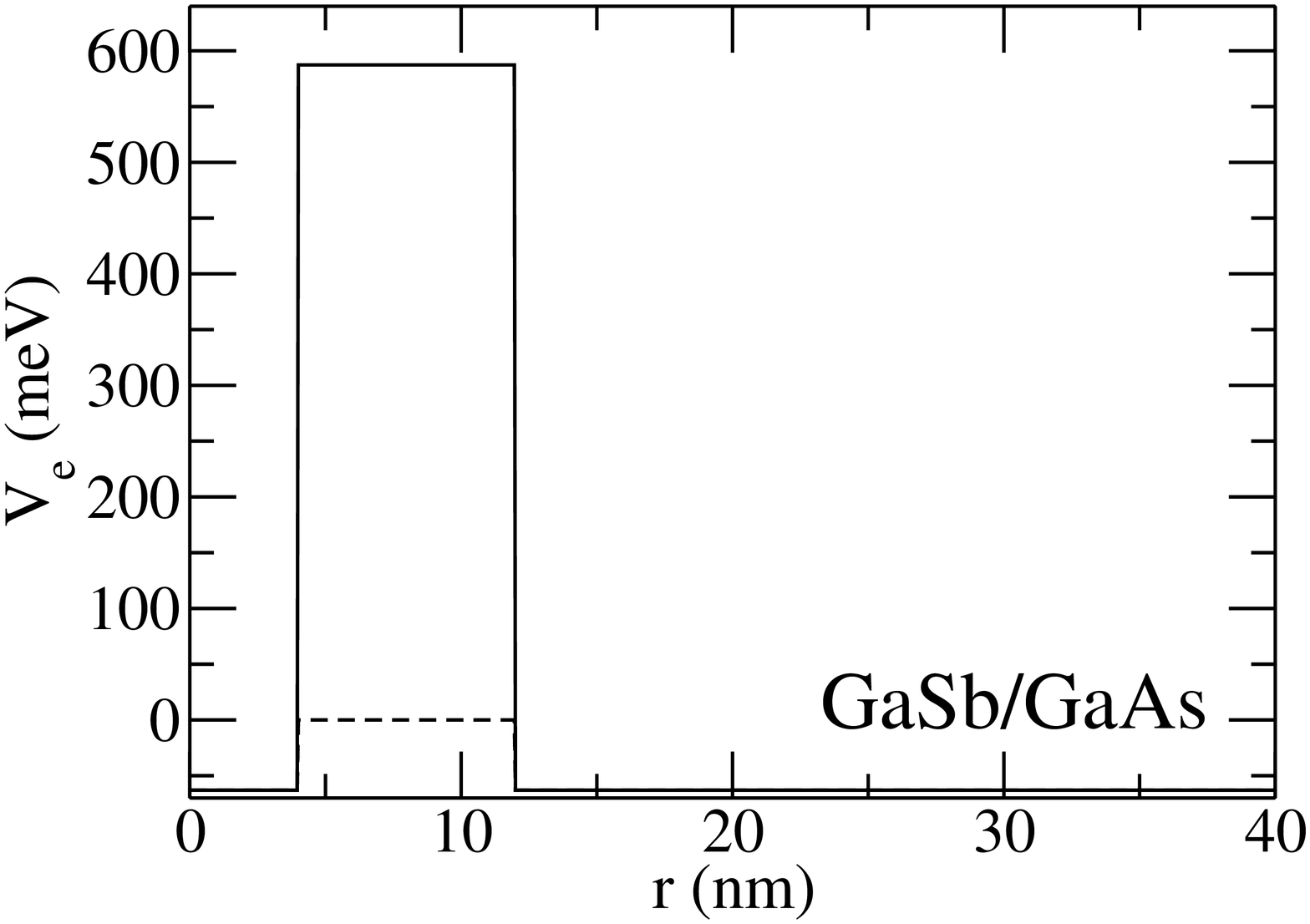, width = 5.9cm, angle = -90}
\caption{The electron potential in the $x$-$y$-plane for $z = 0$ of the GaSb/GaAs nanoring without (dashed) and with (solid) strain included.}
\label{figk}
\end{center}
\end{figure}
}

\newcommand{\taba}
{
\begin{table*}
\caption{Selected material parameters used in the calculation.}
\label{taba}
\begin{ruledtabular}
\begin{tabular}{ccccccc}
 & GaAs\footnotemark[1] & InP\footnotemark[2] & Ga$_{0.51}$In$_{0.49}$P\footnotemark[2] & InAs\footnotemark[1] & Al$_{0.6}$Ga$_{0.4}$Sb\footnotemark[1] & GaSb\footnotemark[1] \\
\hline
$a_0$ (\AA) & 5.65325& 5.8687& 5.65532& 6.0583 & 6.1197& 6.0959\\
$E_g$ (eV) & 1.519& 1.424& 1.97& 0.417& 1.7564& 0.812\\
$m_e$ & 0.067& 0.077& 0.125& 0.026 & 0.0996& 0.039\\
$m_h$\footnotemark[3] & 0.36& 1.67& 0.6 \footnotemark[4] & 0.51& 0.56& 0.71\\
$a_c$ (eV) & -7.17& -7.0& -7.5& -5.08& -5.7& -6.0\\
$a_v$ (eV) & 1.16& 0.4& 0.4& 1.0& 1.16& 1.1\\
$b$  (eV) & -2.0& -2.0& -2.0& -1.8& -1.61& -1.675\\
\end{tabular}
\end{ruledtabular}
\footnotetext[1]{taken from Ref.~\onlinecite{VMR2001}.}
\footnotetext[2]{taken from Ref.~\onlinecite{TPJ+2002}.}
\footnotetext[3]{$m_{hh}^{[110]} = \frac{1}{2}(2 \gamma_1 - \gamma_2 - 3 \gamma_3)$.~\cite{VMR2001}}
\footnotetext[4]{taken from Ref.~\onlinecite{JPP2001}.}
\end{table*}
}

\newcommand{\tabb}
{
\begin{table}
\caption{Chemical band edges Eq. (\ref{CVO}) in meV, relative lattice mismatch $\epsilon_\lambda$ and static dielectric constant $\epsilon_S$.}
\label{tabb}
\begin{ruledtabular}
\begin{tabular}{ccccccc}
 & GaAs/AlGaAs & InP/GaInP & InAs/AlGaSb& GaSb/GaAs \\
\hline
$E_e$ & -257 & -600 & -1673 & 63 \\
$E_h$ & -110 & 50 & 332 & -770 \\
$\epsilon_\lambda$ (\%)&  0 & 3.81 & -1.00 & 7.83 \\ 
$\epsilon_S$ & 12.5 & 12.6 & 15 \footnotemark[1] & 12.5 \\
\end{tabular}
\end{ruledtabular}
\footnotetext[1]{taken from Ref.~\onlinecite{XCQ1992}.}
\end{table}
}

\begin{document}

\title{Optical exciton Aharonov-Bohm effect, persistent current, and magnetization in semiconductor nanorings of type I and II}

\author{M. Grochol, F. Grosse, and R. Zimmermann}

\affiliation{Institut f\"{u}r Physik der Humboldt-Universit\"{a}t zu Berlin, Newtonstr. 15, 12489 Berlin, Germany}

\date{\today}

\begin{abstract}
The optical exciton Aharonov-Bohm effect, i. e. an oscillatory component in the energy of optically active (bright) states, is investigated in nanorings. It is shown that a small effective electron mass, strong confinement of the electron, and high barrier for the hole, achieved e. g. by an InAs nanoring embedded in an AlGaSb quantum well, are favorable for observing the optical exciton Aharonov-Bohm effect. The second derivative of the exciton energy with respect to the magnetic field is utilized to extract Aharonov-Bohm oscillations even for the lowest bright state unambiguously. A connection between the theories for infinitesimal narrow and finite width rings is established. Furthermore, the magnetization is compared to the persistent current, which oscillates periodically with the magnetic field and confirms thus the non-trivial (connected) topology of the wave function in the nanoring.
\end{abstract}


\maketitle

\section{Introduction}
The original Aharonov-Bohm effect (ABE) is found only for charged particles~\cite{AB59} as a purely quantum mechanical effect showing the important role of the vector potential. The ground state energy of a charged particle oscillates with the magnetic flux $\Phi_B =\pi r_0^2 B$ if the particle orbits in a ring around an infinitely long solenoid with radius $r_0$ where the magnetic field $B$ is concentrated. The oscillation period is given in units of the magnetic flux quantum $h / e$. The ABE has been confirmed experimentally e.g. in mesoscopic metal rings,~\cite{ODN+1998} carbon nanotubes,~\cite{BSS+1999} and in doped semiconductor InAs/GaAs nanorings.~\cite{LLG+2000} The persistent current (PC) and the magnetization induced by an electron orbiting in mesoscopic metal~\cite{LDD+1990, CWB+1991} and semiconductor~\cite{MCA1993} ring have been  measured also. Interacting electrons in the ring exhibits both ABE and the PC.~\cite{WF1994, WFC1995}

However, the exciton being a composite particle consisting of electron and hole, has zero total charge. Theoretical studies on the basis of a simplified model (zero width of the nanoring)~\cite{Ch1995, RR2000,MC2004,CM2005,SUS2005, BPS+2006} have demonstrated that the optical exciton Aharonov-Bohm effect (X-ABE), i. e. an oscillatory component in the energy of the optically active (bright) states, exists. However, calculations including the finite width of the rings~\cite{HZLX2001,SU2001, GBW2002} could not confirm these findings for the ground state. A recent calculation on two-dimensional annular lattices~\cite{PDER2005} indicated that the X-ABE of the ground state for nanorings exists, but in this model the energy shift quadratic in the magnetic field was neglected. It is well-known that for a ring radius $r$ much larger than the exciton Bohr radius $a_B$, the X-ABE is practically not observable.~\cite{RR2000} Several proposals have been made to overcome this limitation, such as applying an electric field to separate electron and hole~\cite{MC2003, ZZC2005} or different confinements for electron and hole.~\cite{GUKW2002} The effect of weak disorder or impurity scattering (in general losing the cylindrical symmetry) has been investigated with the result that optically non-active (dark) states can become bright ones.~\cite{SUG2004, SUS2005} In experiment, there are contradictory results: An ensemble of InP/GaAs type II quantum {\it dots} has been studied in Ref.~\onlinecite{RGC2004}. A theoretical explanation based on~Ref.~\onlinecite{KKG1998} indicated some X-ABE oscillations in a single dot. However, in a very recent single dot experiment~\cite{GGN+2006} on InP/GaAs quantum dots (grown under different conditions) no oscillations have been observed. Nevertheless, the X-ABE in {\it nanorings} has not been observed yet. But the ABE has been observed for charged excitons (complex of exciton and electron) in nanoring.~\cite{BKH+2003}

The aim of this work is to calculate the X-ABE, the PC, and the magnetization of the lowest optically active state in nanorings with finite width. A model which captures basic features of real materials, specifically different band alignments and strain, is used to investigate which material parameters are especially favorable for strong X-ABE (the persistent current or the magnetization). However, we stress that it is {\it not} the aim of this paper to model material properties with the most accurate description. Out of this reason some effects which may play an important role in selected materials, like piezoelectric fields, image charge effects, or even valence and conduction band mixing, will be neglected. After describing the theory in Sec. II, the results for the X-ABE are presented in Sec. III, followed by the discussion in Sec. IV. The persistent current and the magnetization are considered in the Sec. V. The paper is summarized in Sec. VI.

\figa

\section{Theory}

\subsection{Exciton Hamiltonian}

Excitons in a nanoring are described here within the envelope function formalism and applying the effective mass approximation (assuming parabolic bands). Including a constant $B$-field perpendicular to the ring plane, the Hamiltonian of a single exciton takes the following form
\begin{eqnarray}
\label{H01}
\nonumber \hat{H} & = &\sum_{a=e,h} \biggl ( \frac{1}{2m_{a, \|}} ( \hat{{\bf p}}_a - q_a {\bf A}({\bf r}_a))^2 + \frac{1}{2m_{a, \perp}} \hat{p}^2_{z_a} \\
\nonumber & & + U_a(z_a) + V_a({\bf r}_a) \biggr ) \\
& & -\frac{e^2}{4\pi \epsilon_0 \epsilon_S \sqrt{({\bf r}_e - {\bf r}_h)^2 + (z_e - z_h)^2}} 
\end{eqnarray}
where $"a"$ denotes either electron (e) or hole (h), $m_{a, \|}$ is the in-plane and $m_{a, \perp}$ is the growth ($z$-) direction carrier effective mass ($\|$ is dropped in the following), $q_a$ is the charge ($q_e = -e$, $q_h = e$), $U_a(z_a)$ is the confinement potential in the growth direction, $V_a({\bf r}_a)$ is the lateral confinement, and $\epsilon_S$ is the static dielectric constant. ${\bf r}_a$ denotes the two-dimensional in-plane coordinates while $z_a$ is the coordinate in the growth direction. The Coulomb symmetric gauge of the vector potential is used: ${\bf A} ({\bf r}) = \frac{1}{2} {\bf B} \times {\bf r}$. The spin degrees of freedom would bring in a term linear in the $B$-field (neglecting spin-orbit coupling) \cite{Ivchenko}
\begin{eqnarray}
\label{Hspin01}
\hat{H}^{spin} & = & \sum_{a=e,h} g^*_a \mu_{B}B \sigma^z_a,
\end{eqnarray}
where $g^*_a$ are effective $g$-factors for electron and hole, $\mu_{B} = e \hbar / 2 m_0$ is the Bohr magneton ($m_0$ being the bare electron mass), and $\sigma^z$ is the Pauli spin matrix. Since electron and hole have different $g$-factors, the spin term does not vanish for the exciton Hamiltonian. \cite{Bla94} This spin dependent part (Zeeman splitting) gives only a linear addition to the total exciton energy. For the sake of simplicity, it is not included in the following analysis. Only heavy hole bright exciton states (total angular momentum $J=1$) are considered in what follows.\par
Assuming that the nanoring is embedded in a narrow quantum well (schematically plotted in Fig.~\ref{figa}), a separation of the wave function for in-plane and growth direction is adopted (single sublevel approximation),
\begin{equation}
\label{WF}
\Phi({\bf r}_e, {\bf r}_h, z_e, z_h) = \Psi({\bf r}_e, {\bf r}_h) v_e(z_e)v_h(z_h),
\end{equation}
where $v_a(z_a)$ are confinement wave functions. Furthermore, cylindrical symmetry is assumed for the lateral confinement. The single-exciton Hamiltonian Eq.~(\ref{H01}) is rewritten in polar coordinates in the following way
\begin{eqnarray}
\label{Ham01}
\nonumber \hat{H} &=& \sum_{a = e,h} \left[  -\frac{\hbar^2}{2m_a} \frac{1}{r_a} \frac{\partial}{\partial r_a} \left (r_a \frac{\partial}{\partial r_a} \right) + \right . \\ 
\nonumber && \left .\frac{1}{2m_a r_a^2} \left ( -i \hbar \frac{\partial}{\partial \phi_a} - q_a \frac{eB}{2} r_a^2 \right )^2 + V_a(r_a) \right]\\
&& +V_C(r_e, r_h, \phi_e-\phi_h)
\end{eqnarray}
introducing the averaged Coulomb potential
\begin{eqnarray}
\label{Coulomb}
V_C(r_e, r_h, \phi) = - \int dz_e dz_h v_e^2(z_e) v_h^2(z_h) \nonumber \\
 \times \frac{e^2}{4\pi \epsilon_0 \epsilon_S \sqrt{r_e^2 + r_h^2 - 2r_e r_h \cos(\phi) + (z_e - z_h)^2}}.
\end{eqnarray} 
Due to the cylindrical symmetry of the Hamiltonian Eq.~(\ref{Ham01}) a transformation to new (Jacobi) angle coordinates is convenient,
\begin{eqnarray}
\phi &=& \phi_e - \phi_h \; , \qquad \Phi = \frac{1}{2} (\phi_e + \phi_h)\\
\frac{\partial}{\partial \phi_e} &=& \frac{\partial}{\partial \phi} + \frac{1}{2} \frac{\partial}{\partial \Phi} \; , \qquad
\frac{\partial}{\partial \phi_h} = -\frac{\partial}{\partial \phi} + \frac{1}{2} \frac{\partial}{\partial \Phi},
\end{eqnarray}
where $\phi$ ($\Phi$) is the relative (average) angle. After transforming Eq.~(\ref{Ham01}), the relation $[\hat{H}, -i\hbar \frac{\partial}{\partial \Phi}] = 0$ can be easily verified. This enables the wave function factorization
\begin{eqnarray}
\label{fact01}
\Psi(r_e, r_h, \phi, \Phi) = \psi(r_e, r_h, \phi) \frac{e^{i m \Phi}}{\sqrt{2 \pi}},
\end{eqnarray} 
where $m$ is an integer quantum number characterizing the total angular momentum of the exciton envelope function. The exciton oscillator strength $f_\alpha$ of the state $\alpha$ is determined as follows~\cite{HaKo} 
\begin{eqnarray}
\label{os01}
f_\alpha = d_{cv} \int d{\bf r} \, \Psi_\alpha({\bf r}, {\bf r}),
\end{eqnarray}
where $d_{cv}$ is the interband dipole matrix element including the state independent contribution of the confinement wave functions $v_a(z_a)$. The absorption spectrum (or optical density) is defined as
\begin{eqnarray}
\label{abs}
D(E) = \sum_\alpha \pi f_\alpha^2 \, \delta(E - E_\alpha).
\end{eqnarray}
Introducing the factorization Eq.~(\ref{fact01}), Eq.~(\ref{os01}) simplifies to
\begin{eqnarray}
\label{os02}
f_\alpha = d_{cv} \sqrt{2 \pi} \delta_{m,0} \int \psi_\alpha(r_e, r_e, 0)  \;r_e dr_e.
\end{eqnarray} 
Therefore, only states with $m = 0$ are optically active. Since the focus of this paper is to calculate {\it optically active} states, we set $m=0$ throughout and end up with a Hamiltonian depending on three coordinates ($r_e$, $r_h$, $\phi$) 
\begin{eqnarray}
\label{Ham02}
\nonumber \hat{H} = && \sum_{a = e,h} \left[ \frac{\hbar^2}{2m_a} \left ( -\frac{1}{r_a} \frac{\partial}{\partial r_a} \left (r_a \frac{\partial}{\partial r_a} \right) \right.  \right.\\
\nonumber && \left. \left.  + \frac{1}{r_a^2} \left ( -i \frac{\partial}{\partial \phi} + \frac{eB}{2 \hbar} r_a^2 \right)^2 \right ) + V_a(r_a) \right] \\
&& + V_C(r_e, r_h, \phi).
\end{eqnarray}
In order to obtain the eigenenergies $E_\alpha$ of the Hamiltonian the wave function is expanded into relative angular momentum eigenstates
\begin{eqnarray}
\label{expan}
\psi_\alpha(r_e, r_h, \phi) = \sum_l u_{l, \alpha}(r_e, r_h)  \frac{e^{il\phi}}{\sqrt{2 \pi}}.
\end{eqnarray}
The Hamilton matrix for the functions $u_l(r_e, r_h)$ can be derived straightforwardly 
\begin{eqnarray}
\label{hamexp}
\nonumber \hat{H}^{ll'}(r_e, r_h) &=& \delta_{ll'} \sum_{a = e,h} \left[  \frac{\hbar^2}{2m_a} \left ( -\frac{1}{r_a} \frac{\partial}{\partial r_a} \left (r_a \frac{\partial}{\partial r_a} \right) \right. \right. \\ 
\nonumber && \left. \left. + \frac{1}{r_a^2} \left ( l + \frac{eB}{2 \hbar} r_a^2 \right)^2 \right ) + V_a(r_a) \right]\\
 && + \tilde{V}_C^{l-l'}(r_e, r_h).
\end{eqnarray}
Kinetic and confinement terms (square brackets) are diagonal in $l$. The Coulomb potential is non-diagonal in $l$ and given by
\begin{eqnarray}
\tilde{V}_C^{k}(r_e, r_h) &=& \frac{1}{2 \pi} \int_0^{2 \pi} V_C(r_e, r_h, \phi) \cos(k \phi) \, d \phi.
\end{eqnarray}
We note that the diamagnetic shift of state $\alpha$ is usually defined in literature as~\cite{GGZb2005}
\begin{eqnarray}
\Delta E_\alpha (B) = E_\alpha (B) - E_\alpha (0).
\end{eqnarray}
Our goal is to extract from the total diamagnetic shift $\Delta E_\alpha (B)$ the {\it oscillatory} component.
  
In the following, the wave function of the state $\alpha$ is analyzed studying the correlated one-particle densities
\begin{eqnarray}
\label{densitye}
n^{(e)}_\alpha(r_{e},\phi) = r_{e}  \int {\textrm d} r_{h} r_{h} |\psi_\alpha(r_e, r_h, \phi)|^2,\\
\label{densityh}
n^{(h)}_\alpha(r_{h},\phi) = r_{h}  \int {\textrm d} r_{e} r_{e} |\psi_\alpha(r_e, r_h, \phi)|^2,
\end{eqnarray}
for which the angular position of the second particle is fixed at say $\phi_a = 0$. Another possibility would be to fix the $r_a$ coordinate instead of integrating over $r_a$. In the case of strong confinement in the ring both approaches are equivalent. The numerical solution is performed by the Lanczos method giving only a few lowest eigenstates of the Hamiltonian Eq.~(\ref{Ham02}). The absorption spectrum Eq.~(\ref{abs}) is calculated by the time propagation method for the optical interband polarization.~\cite{GGZ2005}

\subsection{Persistent current and magnetization}

Even though the exciton is a neutral particle, it can posses a current at a finite $B$-field: electron and hole orbit in the nanoring under the $B$-field in opposite directions, and since they have opposite signs of their charges, their current contributions do add. The exciton persistent current is closely related to the ABE as already pointed out in the literature.~\cite{GUKW2002, MC2004, CM2006} 
The one-particle current density operator at position ${\bf r}$ is defined as
\begin{eqnarray}
\label{cur}
\nonumber \hat{\bf J}_a ({\bf r}) &=& \frac{q_a}{2m_a} \left [ \left ( \hat{\bf p}_a - q_a {\bf A}({\bf r}_a) \right ) \delta ({\bf r} - {\bf r}_a) \right. \\
&& \left. + \delta ({\bf r} - {\bf r}_a) \left ( \hat{\bf p}_a - q_a {\bf A}({\bf r}_a) \right ) \right].
\end{eqnarray}
The expectation value of the radial current $\hat{J}_{a, r}$ is nonzero only for continuum states and will not be discussed further. In the present cylindrical symmetry, the azimuthal current $\hat{J}_{a, \phi}(r, \phi)$ takes the form
\begin{eqnarray}
\nonumber \hat{J}_{a, \phi}(r, \phi) &=& \frac{q_a}{2m_a} \left [ \left (- i\hbar \frac{1}{r_a} \frac{\partial }{\partial \phi_a} - \frac{q_aBr_a}{2} \right ) \delta({\bf r_a} - {\bf r}) \right. \\
&& \left. + \delta({\bf r_a} - {\bf r}) \left (- i\hbar \frac{1}{r_a} \frac{\partial }{\partial \phi_a} - \frac{q_aBr_a}{2} \right ) \right].
\end{eqnarray}
The total exciton current consists of the electron and hole ones, which have to be added and integrated over the cross section of the nanoring,
\begin{eqnarray}
\label{pc1}
I_\alpha (\phi) &=& \int dr \; \langle \alpha |\hat{J}_{e, \phi}(r, \phi) + \hat{J}_{h, \phi}(r, \phi)| \alpha \rangle.
\end{eqnarray}
In accordance with Kirchhoff's laws of current conservation $I_\alpha (\phi)$ is independent of angle $\phi$. 

The one-particle magnetization operator is defined as
\begin{eqnarray}
\hat{\bf M}_a ({\bf r}) = \frac{1}{2} {\bf r} \times \hat{\bf J}_a ({\bf r}).
\end{eqnarray}
The only nonzero expectation value of the magnetization integrated over all space is directed along $z$ and has the magnitude
\begin{eqnarray}
M_\alpha = \pi \int dr \, r^2 \; \langle \alpha |\hat{J}_{e, \phi}(r, \phi) + \hat{J}_{h, \phi}(r, \phi)| \alpha \rangle.
\end{eqnarray}
With the expansion Eq.~(\ref{expan}) the persistent current and the magnetization can be written explicitly as
\begin{eqnarray}
\label{PC2}
\nonumber I_\alpha &=& \frac{e \hbar}{2 \pi} \sum_l  \int dr \, r \left (\frac{l}{r^2} - \frac{1}{2\lambda^2} \right ) \\
&& \times \int dr' r' \left ( \frac{u_{l, \alpha}^2(r, r')}{m_e} + \frac{u_{l, \alpha}^2(r', r)}{m_h} \right ),\\
\label{magnet2}
\nonumber M_\alpha &=& \frac{e \hbar}{2} \sum_l  \int dr \, r \left (l - \frac{r^2}{2\lambda^2} \right ) \\
&& \times \int dr' r' \left ( \frac{u_{l, \alpha}^2(r, r')}{m_e} + \frac{u_{l, \alpha}^2(r', r)}{m_h} \right ).
\end{eqnarray}
Comparing Eq.~(\ref{PC2}) and Eq.~(\ref{magnet2}), the difference is an additional factor $\pi r^2$ in the integrand of the magnetization. The expression for the magnetization can also be derived using the Hellmann-Feynman theorem
\begin{eqnarray}
\label{magnet}
M_\alpha = -\frac{d E_\alpha(B)}{d B}.
\end{eqnarray} 

\subsection{Infinitesimal narrow ring (simplified model)}
In the limiting case of strong electron and hole confinement which can be achieved either by combining materials with appropriate band alignments or by the inclusion of strain,~\cite{GGZ2006} the wave function can be further factorized
\begin{eqnarray}
\label{fact02}
\psi(r_e, r_h, \phi) = f_e(r_e) f_h(r_h) \chi(\phi),
\end{eqnarray} 
where $f_a(r_a)$ are one-particle confinement wave functions in the radial direction centered at $R_a$. The eigenfunction $\chi(\phi)$ can be expanded as in the previous expansion Eq.~(\ref{expan})
\begin{eqnarray}
\label{expan2}
\chi_\alpha(\phi) = \sum_l u_{l, \alpha} \frac{e^{il \phi}}{\sqrt{2 \pi}}.
\end{eqnarray} 
After averaging the Hamiltonian Eq.~(\ref{hamexp}) with functions $f_a(r_a)$ one gets a new Hamilton matrix
\begin{eqnarray}
\label{SilvaModel1}
\nonumber \hat{H}^{ll'} &=& \delta_{ll'} \sum_{a = e,h} \left [ \frac{\hbar^2}{2 m_a R_a^2} \left (l + \frac{e B}{2 \hbar} R_a^2 \right)^2 \right ] \\
&& + \tilde{V}^{l-l'}_C(R_e, R_h)\\
\label{SilvaModel2}
\nonumber &=& \delta_{ll'} \left [ \frac{\hbar^2}{2 \mu_X R_X^2} \left (l + \frac{e B}{2 \hbar} R_X^2 \right)^2 + \Delta E^{(2)} (B) \right ]\\
&& + \tilde{V}^{l-l'}_C(R_e, R_h) ,
\end{eqnarray}
where $\mu_X = m_e m_h / (m_e + m_h)$ is the reduced exciton mass. The energies of the radial motion are omitted. In forming a complete square from the electron and hole kinetic term Eq.~(\ref{SilvaModel2}), an effective ring radius for the exciton 
\begin{eqnarray}
R^2_X = \frac{R_e^2 R_h^2\left(m_e + m_h \right)}{m_e R_e^2 + m_h R_h^2},  
\end{eqnarray}
and a residual energy
\begin{eqnarray}
\label{DS}
\Delta E^{(2)}(B) &=& \frac{e^2 B^2}{8} \frac{(R_e^2 - R_h^2)^2}{m_e R_e^2 + m_h R_h^2},
\end{eqnarray}
appear. 

We note that this Hamiltonian has been intensively studied since its introduction.~\cite{Ch1995} The further simplification of the Coulomb potential to $\tilde{V}^{k}_C(R_e, R_h) = v_C$ (contact potential) enables an analytical solution.~\cite{RR2000, MC2004, CM2005} The case of different electron and hole radii has been studied also.~\cite{GUKW2002, SUG2004, SUS2005, BPS+2006}
 
The total energy can be written as 
\begin{eqnarray}
\label{sep}
E_\alpha (B) = E_\alpha (0) + \Delta E_\alpha^{(1)}(B) + \Delta E_\alpha^{(2)}(B).
\end{eqnarray}
From the structure of Eq.~(\ref{SilvaModel2}) it is clear that $\Delta E_\alpha^{(1)}(B)$ is a strictly periodic function of $B$ with the period~\cite{Ch2002}
\begin{eqnarray}
\label{BP}
B_P &=& \frac{2 \hbar}{e} \frac{1}{R_X^2},
\end{eqnarray}
and is called oscillatory component. 

The persistent current induced by an exciton in state $\alpha$ (Eq.~(\ref{PC2})) can be also obtained from a version of the Hellman-Feynman theorem after introducing the one-particle flux~\cite{WFK1994, MC2004} $\Phi_{B,a} = \pi R_a^2 B$
\begin{eqnarray}
\label{percur}
I_\alpha = -\frac{\partial E_\alpha(\Phi_{B,e}, \Phi_{B,h})}{\partial \Phi_{B,e}} - \frac{\partial E_\alpha(\Phi_{B,e}, \Phi_{B,h})}{\partial \Phi_{B,h}}.
\end{eqnarray}
In the present case, however, this would call for a calculation of the problem in dependence on two different (fictitious) $B$-fields in $E_\alpha(\Phi_{B,e}, \Phi_{B,h})$. From the Hamiltonian Eq.~(\ref{SilvaModel2}) the exciton flux can be defined as
\begin{eqnarray}
\Phi_{B, X} = \pi R_X^2 B
\end{eqnarray}
and the exciton energy can be regarded as a function of the exciton flux $E_\alpha(\Phi_{B,X})$. It turns out that the exciton PC calculated from the definition Eq.~(\ref{percur}) is equivalent to the derivative of the oscillatory component of the exciton energy only,
\begin{eqnarray}
\label{percur2}
I_\alpha = - \frac{1}{\pi R_X^2} \frac{d \Delta E_\alpha^{(1)} (B) }{d B}.
\end{eqnarray}
This means that the term quadratic in $B$ ($\Delta E_\alpha^{(2)}$) does not contribute to the PC. From the general definition of the magnetization Eq.~(\ref{magnet}) a relation between the magnetization and the PC can be found easily,
\begin{eqnarray}
\label{magnetpc}
M_\alpha = \pi R_X^2 I_\alpha -\frac{d \Delta E_\alpha^{(2)} (B)}{d B}.
\end{eqnarray}
The second term is the intrinsic magnetization originating from the inner electron and hole motion in the exciton while the first one being proportional to the PC, is related to the non-trivial (connected) topology of the wave function. Only in the case of identical electron and hole radii, $R_e = R_h$, the PC and the magnetization are proportional, since the term $\Delta E_\alpha^{(2)}$ is absent.

In the following, we will refer to finite width (zero width) ring equivalently as to full (simplified) model.

\subsection{Strain}
The strain plays an important role in some materials and should be included at least in a first approximation. Recently the simple but for our purpose well adjusted approach~\cite{D1998, TPJ+2002, JPP2003} of isotropic elasticity in nanostructures has been developed. Assuming barrier and well material to be identical in their elastic properties, the strain tensor $\epsilon_{ij}$ reads 
\begin{eqnarray}
\nonumber \epsilon_{ij} ({\bf r}) = -\epsilon_\lambda \theta({\bf r} \in {\rm Ring}) - \frac{\epsilon_\lambda}{4 \pi}
\frac{1 + \nu}{1 - \nu} \oint_{S'} \frac{(x_i - x_i') dS_j'}{|{\bf r} - {\bf r}'|^3}\\
\end{eqnarray}
where $S'$ is the nanoring surface over which is integrated, $\nu$ is the Poisson ratio, $\epsilon_\lambda = (\lambda_{\mbox{Ring}} - \lambda_{\mbox{Well}}) / \lambda_{\mbox{Well}}$ is the relative lattice mismatch between ring lattice constant $\lambda_{\mbox{Ring}}$ and well lattice constant $\lambda_{\mbox{Well}}$. The electron and heavy hole band edges are modified due to the strain in the simplest form~\cite{JPP2003} 
\begin{eqnarray} 
\nonumber V_e({\bf r}) &=& a_c tr({\bf \epsilon})({\bf r}) + E_e \theta({\bf r} \in {\rm Ring}) \; ,\\
\nonumber V_h({\bf r}) &=& -a_v tr({\bf \epsilon})({\bf r}) - b [(\epsilon_{xx}({\bf r})
  + \epsilon_{yy}({\bf r})) / 2 - \epsilon_{zz}({\bf r})]\\
   && + E_h \theta({\bf r} \in {\rm Ring})
\label{CVO}
\end{eqnarray}
where $a_c$ ($a_v$, $b$) are conduction (valence) band deformation potentials and
$E_{e(h)}$ strain-free band edge discontinuities ($E_a < 0$ are confining potential, the barrier values are set to zero). The strain contribution is included in the Hamiltonian Eq.~(\ref{Ham02}) after being averaged over $z$ in the same way as the Coulomb potential Eq.~(\ref{Coulomb}). This approximation holds well due to the strong confinement in the $z$ direction.

\taba
\tabb

\section{Results of X-ABE}

Here, we present the results of X-ABE for all material and confinement types. 

We investigate: (i) Type I nanoring, where the electron and the hole are confined together, (ii) type II-A nanoring, where the electron is confined in the ring and the hole has a ring-like barrier, and (iii) type II-B nanoring, where the hole is confined in the ring and the electron has a ring-like barrier (schematically shown in Fig.~\ref{figa}). The  well and ring material parameters are summarized in Tab.~\ref{taba} and~\ref{tabb}. The effective masses are chosen according to the material in which the particle is found predominantly. 

In the investigation of the X-ABE we concentrate on the optically active state with the lowest eigenenergy $E_\alpha$ in what follows, since its proper confirmation in finite width nanorings represents an open question. We propose a new method for observing oscillations: The second derivative of the energy with respect to the $B$-field. It has already been mentioned in the introduction and is shown below that the oscillations of the excited states are much easier visible due to their higher kinetic energy leading to much larger exciton extension. 

We have chosen $B$-field strengths up to $B=25$~T which can be easily achieved in experiment.

\figj

\figc

\subsubsection{Type I ring - GaAs/AlGaAs}

The choice of GaAs/AlGaAs for type I structure is rather straightforward since it is the most frequently investigated direct semiconductor. The strain can be safely neglected in this structure due to the small lattice mismatch (in contrast to previously investigated self-assembled InAs/GaAs nanorings~\cite{LLG+2000, BWR2004}). There is even a newly developed technique which allows to grow concentric nanorings.~\cite{KMO+2005} The structure investigated here consists of a Al$_{0.23}$Ga$_{0.77}$As 4~nm wide quantum well between Al$_{0.3}$Ga$_{0.7}$As barriers. A nanoring of pure GaAs is placed inside the Al$_{0.23}$Ga$_{0.77}$As well.

As an example we discuss the absorption spectrum Eq.~(\ref{abs}) of an GaAs/AlGaAs nanoring as plotted in Fig.~\ref{figj}a where the oscillations of the excited states are indeed clearly visible, while the ground state shows only a smooth and monotonic energy shift. Before discussing the properties of the lowest bright state we focus our attention to one interesting feature of excited states, namely the anti-crossing marked by a circle in Fig.~\ref{figj}. On a first glance, the absorption spectrum resembles the result of the simplified model~\cite{CM2005}: The first three lowest bright states can also be found in the simplified model as plotted in~Fig.~\ref{figj}b-d (being even, even and 
odd with respect to $\phi$ at $B=0$~T). Only the fourth state (Fig.~\ref{figj}e) cannot be found in the simplified model since the hole sits in its first excited radial state, which is absent in a zero width ring. Its overlap with the electron part and consequently the oscillator strength is, however, tiny. Nevertheless, this even state manifests itself strongly by the anti-crossing with the next even state at around $B=13$~T. This kind of anticrossing, even though somewhat marginal, goes beyond the description of the simplified model. From now on, let's concentrate on the lowest bright state.

Up to now there has not been any clear evidence of oscillations of the lowest bright (ground) state for finite width nanorings. The problem becomes evident looking at Fig.~\ref{figc}a, where on the first glance the only dependence of the energy on the $B$-field is the smooth and monotonic increase. Although for nanorings of finite width a separation of the diamagnetic shift like Eq.~(\ref{sep}) is not possible in a strict sense, we will understand in the following $\Delta E_\alpha^{(1)}$ as the oscillating part and $\Delta E_\alpha^{(2)}$ as the smooth monotonic part. The behavior of the exciton ground state energy in the limit $B \to 0$ has been studied in Ref.~\onlinecite{WR1998} finding a non-trivial dependence on the one-particle confinement and exciton relative motion. In the present case, the strong electron (hole) ring confinement fixes the electron (hole) radial position $r_{e(h)}$ and the strong Coulomb interaction  fixes the relative distance $r$, which means that the quadratic dependence on $B$  and consequently its contribution to the second derivative are almost constant with the $B$-field. This enables to extract the second - oscillatory - component from the second derivative as seen in Fig.~\ref{figc}b. The strong dependence of the oscillations amplitude on the ring radius is remarkable. The expectation values of $R_a^2 = \langle r^2_ {e(h)}\rangle$ at $B = 0$~T from the full solution were used as input parameters in the simplified model. A comparison of the simplified with the full model shows good agreement for the period and the amplitude of the oscillations, and its strong dependence on the ring radius as well. The main difference is the absence of the term $\Delta E_\alpha^{(2)}$ in the simplified model. It is the finite radial extension of the exciton relative wave function (plotted in Fig.~\ref{figc}c) which gives a nonzero contribution to this energy. After having compared results of both models in this case, further we will discuss only the full solution. Note, however, that the period Eq.~(\ref{BP}) gives generally a good estimate.

Let us direct our attention to type II systems where more pronounced effects are expected.

\figd

\figb

\subsubsection{Type II-A -- InP/GaInP}

InP/GaInP self-assembled quantum dots are of type II-A, and have been investigated since many years both theoretically and experimentally (see~Ref.~\onlinecite{PHJ+2005} and references therein). Possibly, rings may be grown as well, e. g. using the same procedure as for InAs/GaAs~\cite{GMS+1997} or GaAs/AlGaAs~\cite{KMO+2005} nanorings. We have investigated a structure consisting of a 4~nm wide Ga$_{0.51}$In$_{0.49}$P quantum well between AlAs barriers. The nanoring is pure InP. Such a structure guarantees that the hole is always found around the ring (in the  $xy$-plane) and not above or below the ring (in growth ($z$-) direction). This is not a necessary condition for the X-ABE. The situation where the hole (electron) is found above or below is also of interest. This goes beyond the scope of this paper since we would not be able to take advantage of the $z$-separation. The strain plays an important role in this material,~\cite{JPP2003} as is clearly shown in Fig.~\ref{figb}a. 

Since the deep minimum of the hole potential is formed at the inner edge  of the ring (Fig.~\ref{figb}a), the hole is found there. Such a state is named {\it hole-in} (depicted in Fig.~\ref{figd}c). The effective electron-hole separation is thus decreased with respect to the strain-free case. The state hole-in is the ground state for any ring radius. Excluding composition changes, the height of the ring-like barrier for the hole can decrease by changing the well width in the $z$ direction. For high $B$-fields, a transition from type II to type I may occur due the enhanced penetration of the hole wave function into the ring. This has been already predicted for quantum dots in Ref.~\onlinecite{JPP2003}. The energy of the lowest bright states as a function of $B$-field is plotted in Fig.~\ref{figd}a. Again, no evidence of oscillations is seen by the naked eye. The analysis of the second derivative (Fig.~\ref{figd}b) reveals that (i) the amplitude of the oscillation is increased compared to type I (as expected), and (ii) the period of the oscillation is increased as well since the hole samples a smaller magnetic flux (see Eq.~(\ref{BP})) compared to type I. 

\figf

\subsubsection{Type II-A -- InAs/AlGaSb}

The InAs/AlGaSb system has several advantages for observing the X-ABE in type II-A systems compared to InP/GaInP as will be discussed below. This system is less known compared to GaAs/AlAs or InP/GaInP, but as a quantum well structure well-understood and used (see~Ref.~\onlinecite{OFO2005} and references therein). Recently, InAs quantum dots on AlGaSb substrate have been grown.~\cite{YAG+2004} The fact that Al$_{0.6}$Ga$_{0.4}$Sb is an indirect semiconductor~\cite{AJJ+1983} is of less importance since the electron is found predominantly in InAs, which means that the approximation taking into account only the $\Gamma$ point is sufficient. A problem is the small difference (0.083~eV) between the conduction band edge in InAs and the valence band edge in Al$_{0.6}$Ga$_{0.4}$Sb.~\cite{VMR2001} The applicability of the effective mass approximation is questionable here.~\cite{CTC+1996} Nevertheless, we believe as a first approximation~\cite{XCQ1992} it can be adopted. The investigated structure consists of a Al$_{0.6}$Ga$_{0.4}$Sb 4~nm wide quantum well between AlSb barriers. A InAs nanoring is placed in the Al$_{0.6}$Ga$_{0.4}$Sb well. Even though the lattice mismatch between InAs and AlGaSb is small (1\%), our calculation includes strain (see Fig.~\ref{figb}b). 

The influence of the strain on the hole potential for InAs/AlGaSb is shown in Fig.~\ref{figb}b. Compared to InP/GaInP, there are striking differences. The effect of strain is much smaller due to the much smaller lattice mismatch and the sign of the strain contribution is opposite. Instead of compression in the ring as for InP/GaInP (enlarging the bandgap), there is dilatation in the case of InAs/AlGaSb which lowers the bandgap. This leads to the repulsion of the hole from the ring and thus to a weakening of the Coulomb interaction. In contrast to InP/GaInP, the minimum of the hole potential is found outside of the ring (Fig.~\ref{figb}b) for any ring radius. If the hole is found outside the ring, such state is called {\it hole-out} (depicted in Fig.~\ref{figf}c). The difference between the potential value in the middle and outside the ring decreases with increasing inner ring radius. 

As stated above, due to its material properties a large oscillation amplitude is found here, as seen in Fig.~\ref{figf}a and b. Even without any further analysis, the lowest bright state, hole-out, shows clear oscillations. Please note a change of the scale by {\it a factor of ten} in Fig.~\ref{figf}b compared to Fig.~\ref{figd}b! In both cases kinks in $E(B)$ (sharp minima in $d^2E/dB^2$) resemble the one-particle ABE and are consequence of the weak Coulomb interaction. 

\figg

An interesting new effect is found in larger rings, namely a transition from hole-in to hole-out. Depending on the ring geometry one of them is the lowest bright state and the other one the second lowest. The strain profile favors the state hole-out. On the other hand, the Coulomb interaction prefers the state hole-in. As the strain profile in the middle and outside of the ring becomes similar for larger rings, the Coulomb interaction dominates and the state hole-in becomes the lowest bright state. This situation is demonstrated in Fig.~\ref{figg}b.  The lowest bright state changes with increasing $B$-field: from hole-in (Fig.~\ref{figg}b) to hole-out (Fig.~\ref{figg}c and d). The state hole-in has a larger energy shift $\Delta E^{(2)}$, which can be verified by calculating the effective hole radii $\langle r_h^2 \rangle$ and checking Eq.~(\ref{DS}). The transition occurs at around $B=1.5$~T. Since we always follow the lowest bright state the second derivative shows a sharp peak at the transition (Fig.~\ref{figg}a). We note that the small overlap of the hole-in and hole-out wave functions does not allow to distinguish between level crossing and anti-crossing, at least within our numerical precision. Due to the large radius of the ring, the oscillation period is small (according to Eq.~(\ref{BP}) $B_P = 1.9$~T). The decay of the oscillation amplitude is due to a decrease in exciton Bohr radius with $B$-field (compare Fig.~\ref{figg}c and Fig.~\ref{figg}d). 

\figk
\figh

\subsubsection{Type II-B -- GaSb/GaAs}

GaSb/GaAs self-assembled quantum dots of type II-B have attracted a certain interest recently (see~Ref.~\onlinecite{TEL+2004} and references therein). The strain plays a very important role in these structures and modifies significantly the conduction and valence band energies: The strain-free offsets (Eq.~(\ref{CVO})) $E_e = 0.063$~eV and $E_h = -0.77$~eV are modified to $E_e = 0.65$~eV and $E_h = -0.86$ (minimum). The substantial change of the electron potential is shown in Fig.~\ref{figk}. These results are comparable with those in~Ref.~\onlinecite{PP2005}. The investigated structure consists of a GaAs 4~nm wide quantum well between Al$_{0.3}$Ga$_{0.7}$As barriers, a GaSb nanoring is placed in the well. 

The increase of the lowest bright state energy by 40 meV from $B=0$~T to $B=25$~T (Fig.~\ref{figh}a) is large compared to all previous values and again no clue of oscillation is seen. In the second derivative (Fig.~\ref{figh}b), a sharp initial decay is revealed. The origin of this decay can be understood studying the wave function. The hole is strongly confined in this system and the electron potential has a high ring-like barrier as mentioned above. The correlated electron density plotted in Fig.~\ref{figh}c shows that the shallow Coulomb potential localizes the electron part of the wave functions only weakly. The electron is very sensitive to the $B$-field. Due to the quadratic term $\Delta E^{(2)}$, the electron is forced to move quickly towards the hole (i. e. to the nanoring) with increasing $B$-field. This behavior leads to the initial decay in the second derivative (Fig.~\ref{figh}b). Lateron, the wave function stabilizes and oscillations appear (Fig.~\ref{figh}b). Their amplitude is comparable to values found in InP/GaInP. The advantage of the GaSb/GaAs system is the large Bohr radius due to the large electron-hole separation. On the contrary, the disadvantage is the sensitivity of the electron to the $B$-field due to its small mass and shallow confinement.

\section{Discussion of X-ABE}

Now, we compare and discuss the results of the previous section and conclude which material combination is preferential for X-ABE.

Our results show unambiguously that a weakening of the Coulomb interaction increases the oscillation amplitudes, which was already been shown for the simplified model.~\cite{GUKW2002, SUS2005} The reason is that electron and hole can sample the entire ring more easily if the exciton is weakly bound (larger exciton Bohr radius), and the wave function can acquire the necessary ring topology. The mutual confinement of electron and hole (type I) has turned out to be inferior to the systems where electron and hole are separated by the conduction and valence band alignments in the $xy$-plane (type II). One unwanted consequence of the spatial electron-hole separation is that the lowest bright state is not any more the ground state for larger $B$-fields~\cite{Ch2002, GUKW2002, BPS+2006} (in contrast to type I). This may result~e.~g. in losses of photoluminescence intensity if some non-radiative decay channels are present. A calculation of these kinetic effects goes, however, beyond the scope of this paper. 

Comparing the results for different material systems, we find that for large amplitude of X-ABE oscillations the ideal structure is of type II-A with the following properties: Light electron mass, strong electron confinement, and high barrier for holes. These criteria can be discussed qualitatively: (i) The light electron mass leads to a larger Bohr radius and higher probability of the particle to sample the whole nanoring. (ii) Strong electron confinement is needed in order to force the light electron to orbit around the ring. (iii) A high barrier for holes is necessary for a "good" type II nanoring, thus avoiding the penetration of the wave function into the ring. The necessity of a high (deep) potential of the lighter particle has already been pointed out.~\cite{GGZ2006} The material which matches these criteria best is InAs/AlGaSb in our case. This system clearly deserves further investigations, both theoretically and experimentally. 

\figi

\section{Persistent current and magnetization}

After having examined the X-ABE in different materials, we investigate now the persistent currents and the magnetization in a more compact way. 

The PC Eq.~(\ref{PC2}) and the magnetization Eq.~(\ref{magnet2}) can be measured under special conditions: (i) The exciton should be excited into the optically active state (in our case always the lowest one). (ii) The excitation power should be sufficient in order to give a measurable signal but small enough to avoid exciton-exciton interaction. We assume one exciton per nanoring in the following, which corresponds to extremely strong excitation. A more realistic value of the excitation would reduce the scales in Fig.~\ref{figi} accordingly. The state-of-the-art experimental technique (SQUID) enables to measure the magnetization directly.~\cite{MCA1993} The measurement of the current requires additional contacts on the nanoring which may complicate the already difficult assembling of the nanorings even more. A measurement of either the persistent current or the magnetization in the nanoring is an extremely challenging task.

First, the magnetization of the exciton in the ring with finite width (Fig.~\ref{figi}a-d) can be divided roughly also into two contributions according to the analogy with the simplified model Eq.~(\ref{magnetpc}): The first (oscillatory) part comes from the orbiting of electron and hole around the nanoring while the second monotonic part (nearly linear in $B$) is related to the inner exciton motion where electron and hole orbit around each other. The weight of each contribution depends on the wave function topology. The dominance of the first contribution is seen  only in the case of weak Coulomb interaction where the magnetization has even negative values, i. e. pointing into the opposite direction of the $B$-field itself. Such an effect could be interesting for further applications, namely a sign switch of the optically induced coherent magnetization by the $B$-field. Unfortunately, this effect is rather weak. In all other cases both parts contribute with different weights. The linear component is found also for excitons in quantum wells or dots. 

Second, the exciton PC shown in Fig.~\ref{figi}e-h exhibit periodic oscillations for each ring geometry and for all materials, even though the oscillation amplitude may be very small (as e. g. in Fig.~\ref{figi}d). The period and the relative amplitude of the oscillations agree well with those of the second derivative of the energy with respect to the magnetic field. Furthermore, the results confirm the relation Eq.~(\ref{percur2}) qualitatively also for finite width nanorings since they indeed remind of the first derivative of the oscillatory component of the exciton energy. A quantitative comparison is in general not possible since the smooth (non-oscillatory) component of the diamagnetic shift cannot be extracted unambiguously. From the theoretical point of view it turns out that a measurements of the PC would give a more direct information on the  non-trivial ring topology of the wave function.

\section{Summary}

In summary, we have investigated the optical exciton Aharonov-Bohm effect (X-ABE), persistent current and magnetization in nanorings. We have discussed the differences between a simplified and the full model. In the simplified model, the smooth monotonic part of the energy is small or even missing. We have focused on the observability of the X-ABE in type I and II nanorings which can be improved by taking appropriate materials, at least, by one order of magnitude. We have discussed in detail GaAs/AlGaAs (type I), InP/GaInP and InAs/AlGaSb (type II-A), and GaSb/GaAs (type II-B) nanorings and found that also in large InAs/AlGaSb nanorings the X-ABE can be observed. We have shown that a hole-in-hole-out transition occurs in these nanorings. The persistent current is found to be proportional to the first derivative of the oscillatory component of the exciton energy and thus has a close relation to the X-ABE. In the case of the magnetization we have demonstrated that oscillations are superimposed on the linear component related to the inner exciton motion and being independent on the ring topology of the wave function.

\section{Acknowledgment}
We acknowledge fruitful discussions with D. S. Citrin, E. A. Muljarov, and L. Wendler. M. G. acknowledges financial support from the Graduate school Nr. 1025 of the Deutsche Forschungsgemeinschaft.

\bibliographystyle{apsrev}
\end{document}